\documentclass[fleqn,twoside]{article}
\usepackage[headings]{espcrc2}
\usepackage{amsthm,amssymb}

\readRCS
$Id: espcrc2.tex,v 1.2 2004/02/24 11:22:11 spepping Exp $
\usepackage{graphicx, balance}
\usepackage[figuresright]{rotating}

\newcommand{\AmS}{{\protect\the\textfont2
  A\kern-.1667em\lower.5ex\hbox{M}\kern-.125emS}}

\title{\textbf{Two Stage Prediction Process with Gradient Descent Methods Aligning with the Data Privacy Preservation}}

\author{S kumarasawamy\address[DCSE]{Department of Computer Science and Engineering, University Visvesvaraya College of\\~ Engineering, Bangalore University, Bangalore 560 001 India, Contact: kumar.aruna@gmail.com.\\},
{Srikanth P L\addressmark},
{Manjula S H\addressmark},
{K R Venugopal \addressmark},
L M Patnaik\address{Honorary Professor, Indian Institute of Science, Bangalore.}}

\runtitle{Gradient Descent Methods Aligning with the Data Privacy Preservation}
\runauthor{Kumaraswamy S, et al.,}

\begin{document}
\begin{abstract}
Privacy preservation emphasize on authorization of data, which signifies that data should be accessed only by authorized users. Ensuring the privacy of data is considered as one of the challenging task in data management. The generalization of data with varying concept hierarchies seems to be interesting solution. This paper proposes two stage prediction processes on privacy preserved data. The privacy is preserved using generalization and betraying other communicating parties by disguising generalized data which adds another level of privacy. The generalization with betraying is performed in first stage to define the knowledge or hypothesis and which is further optimized using gradient descent method in second stage prediction for accurate prediction of data. The experiment carried with both batch and stochastic gradient methods and it is shown that bulk operation performed by batch takes long time and more iterations than stochastic to give more accurate solution.  
\end{abstract}

\maketitle

\section{INTRODUCTION}
Data mining is a process of extracting the knowledge from large set of data. The knowledge extraction defines a model or rules for performing the accurate analysis on the future data. The initial data set is referred as training data, since it is used as reference for arriving at knowledge.
The analysis performed on the data should not reveal the data as such and hence preserving the privacy of data becomes significant. Intuitively, differential privacy ensures that the system behaves the essentially same way, independent of whether any individual, or small group of individuals, opts in to or opts out of the database [1].  Generalization and suppression are two predominant techniques to achieve the privacy of data [2-3]. Privacy of data is very important in certain domains like hospital, analysis of psychological behaviour of patients [4]. The solution presented in this paper is to apply gradient descent methods on the privacy preserved data. The gradient descent is first order optimization algorithm to find local minimum of the function [5].
\subsection{Motivation}
Gradient descent is a widely used paradigm for solving many optimization problems.  In machine learning or data mining, this optimization function corresponds to a decision model that is to be discovered [6]. Shuguo Han {\it et. al.,} has proposed the solution for application of gradient descent methods on the privacy preserved data. The data is either vertically or horizontally partitioned across communicating parties. The factor for partition is mutually synchronized between each other. In vertical partitioning every communicating party has same set of objects with varying attributes. In this scenario the prediction is first performed on unknown attributes before performing prediction for specific object. In horizontal partitioning every communicating party has disjoin set of records with same set of attributes. The prediction in this scenario is performed on unknown objects. To achieve required accuracy in prediction gradient descents are used in the context of machine learning. The gradient descent follows the iterative approach to minimize the prediction function until local minimum is reached. The local minimum is the threshold which defines the required accuracy factor. The gradient descents are applied on matrix of numbers and the values of unknown columns are determined using random distribution before prediction.
\subsection{Contribution}
The solution proposed works in 2 different stages. First stage defines the model or rules representing the knowledge about the data. The model is defined separately by every communication party. While defining the model the values of unknown attributes are inferred from bamboozled abstract information. The abstract information indicates the high level generalized information. The abstract information is further disguised using disguising factor to 	make other communication party deceived by the information shared and hence referred as bamboozle process. The bamboozling is performed to preserve the privacy at maximum extent.
\vskip 2mm
The data obtained after bamboozling is misleading Meta data. The Meta data is processed by every communication party to find the values of unknown attributes. The aggregation of both unknown and known attributes produces the regression model. The regression model represents the upper bound values which are optimized with the application gradient methods until the required accuracy is reached. The prediction function used here is the quadratic function defined as $w^2F$  where $w$ represents the weight vector used as coefficient of function which is diminished in every iteration of gradient descent until local optimum is reached. The factor that decides the local minimum function is also referred as learning stoppage as it terminates the 	prediction process. There are two approaches in gradient descent $i.e.,$ batch and stochastic.
\vskip 2mm
In stochastic model prediction is performed by considering one data sample at a time. In batch gradient all the samples are processed before updating individual sample. Batch gradient takes longer in its inner loop (large sum) but it can use a larger step size because of this sum. Although these methods assume un-thresholded linear units, they can be easily modified to work on regular perceptions [7-12]. In this paper the behaviour of both batch and stochastic gradient methods are analyzed w.r.t number of iterations and execution time required to perform the prediction process. The bamboozled information is conveyed using standard format which is machine readable. One such Meta data is Resource Description Framework (RDF), in which the information is represented as sequence of triplets. Every RDF [13] should adhered to respective domain Ontology. Ontology [14] defines the concepts and relations among the concepts, RDF describes the web document in the form of triplets. Every RDF triplet is a composition of Subject, Predicate and Object [15]. The RDF is processed by every communicating party to get the values of unknown attributes.
\subsection{Organization}
This section describes the rest of the paper in brief. Section 3 describes the background work and its comparison with current solution. Section 2 describes the other significant related work performed in this area. Section 4 defines the problem. Section 5 defines mathematical model. Section 6 describes the algorithms used to arrive at the solution Section 7 gives the overall system architecture along with the ontology model used in this paper. Section 8 explains the experimental results and data set used to simulate the process. The paper concludes by mentioning the enhancement that can be incorporated in prediction process along with the suppression and the list of references considered by the authors.
\section{RELATED WORK}
Some predominant research is performed in the area of data privacy preservation. Kanishka Bhaduri {\it et. al.,} [16], suggested an approach to allow a user to control the amount of privacy by varying the degree of nonlinearity. It is also shown how the general transformation can be used for anomaly detection in practice for two specific problem instances: a linear model and a popular nonlinear model using the sigmoid function. There is also an analysis on the proposed nonlinear transformation in full generality and then show that, for specific cases, it is distance preserving. A main contribution of this paper is the discussion between the invertibility of a transformation and privacy preservation and the application of these techniques to outlier detection.
\vskip 2mm
 Benjamin C M Fung {\it et. al.,} [17]done the study on the resolution of  a privacy problem in a real-life mashup application for the online advertising industry in social networks, and propose a service-oriented architecture along with a privacy-preserving data mashup algorithm to address the aforementioned challenges. Experiments on real-life data suggest that our proposed architecture and algorithm is effective for simultaneously preserving both privacy and information utility on the mashup data. To the best of our knowledge, this is the first work that integrates high-dimensional data for mashup service.
\vskip 2mm
 Jung Yeon Hwang {\it et. al.,} [18]has presented a short group signature scheme for dynamic membership with controllable linkability.The controllable linkability enables an entity who possesses a special linking key to check if two signatures are from the same signer while preserving anonymity. It can be used for various anonymity-based applications that require necessarily the linkability such as vehicular ad hoc network, and privacy-preserving data mining. Our scheme is sufficiently efficient and so, well-suited for real-time applications even with restricted resources, such as vehicular ad hoc network and Trusted Platform Module.
\vskip 2mm
 ZHU Yu-quan, TANG Yang {\it et. al.,}[19]have addressed  the problem that the existing protocol of secure two-party vector dot product computation has the low efficiency and may disclose the privacy data, a method which is effective to find frequent item sets on vertically distributed data is put forward. The method uses semi-honest third party to participate in the calculation, put the converted data of the parties to a third party to calculate.  Alberto Trombetta {\it et. al.,} [20] have come up with two protocols solving k-anonymity problem on suppression-based and generalization-based $k$-anonymous and confidential databases. The protocols rely on well-known cryptographic assumptions and we provide theoretical analyses to proof their soundness and experimental results to illustrate their efficiency.
\section{BACKGROUND WORK}
 Shuguo Han {\it et. al.,} [6] have proposed the application of stochastic gradient descent methods on the vertically partitioned. The experiment is performed on the matrix of numbers and the data of unknown attributes are generated by using random distribution. But the current solution gives the relevant information to the other communication party to find the values of unknown attributes. Even though the information is relevant it preserves the privacy of data by exposing the upper bound generalized values. To energize the privacy the generalized data is disguised and hence the complete data given other communicating party forms the semantic metadata. Even though gradient descents take more iterations because the of bamboozling process but the accurate prediction is ensured by using descent learning stoppage.

\section{PROBLEM DEFINITION}
Given the database distributed across the communicating parties using vertical partitioning, the main problem is to perform the prediction of single data with proper synchronization but not disclosing each other’s data in other words the privacy of data should not be affected. The prediction is performed using gradient descent methods incorporated as iterative process. The gradient descents are required to applied on privacy preserved data.
\vskip 2mm
{\it Assumptions:} It is assumed that the data is partitioned and both communicating parties should adhered to common ontology model.
\section{MATHEMATICAL MODEL}
\subsection{List of notations used}
Following table enumerates the list of notations used and their purpose while defining the model.
\begin{table}[ht!]
\begin{center}
\caption{Basic Notations}
\begin{tabular} {|l|l|}
\hline
\textbf{Notations}  &  \textbf{Meaning} \\\hline
${\eta}_{s}$ & learning rate for \\
          & Stochastic gradient. \\\hline
${\eta}_{b}$ & learning rate of \\
          &  Batch gradient.\\\hline
$\lambda$ & Learning stoppage \\\hline
$f^{\rightarrow}$ &  Result Vector after\\
          &first stage prediction.\\\hline
$p^{\rightarrow}$ & Result Vector after\\
                  & second stage prediction.\\\hline
$w^{\rightarrow}$ & Weight vector used\\
                  & as a reduction factor.\\\hline
$F({w^{\rightarrow},f^{\rightarrow}})$ & Second stage prediction\\
                                       &function defined as the \\
                                       & function of weight vector\\
                                       &and first stage result vector.\\\hline
$x$ & Un partitioned Data Records.\\\hline
$n$ & Number of Data records.\\\hline
$m$ & Number of attributes\\
    & of every data record.\\\hline
$E^{\rightarrow}$ & Expected Vector\\
                  &for all data records.\\\hline
\end{tabular}
\end{center}
\end{table}
\subsection{Definitions}
{\bf Learning Rate}: The learning rate is factor by which the value is minimized. The learning rate and iterations to reach local minimum is inversely proportional to each other $i.e.,$ If the learning rate is high then the number of iterations required to reach local minimum for a function is less and vice versa.
\vskip 2mm
{\bf \textbf{${\eta}_{s}$ }}- is the Learning rate of stochastic gradient and for the employee case study it is maintained as 0.00001 as the sample values are in hundred to thousand range.
\vskip 2mm
\textbf{${\eta}_{b}$ - } is the Learning rate of batch gradient method and for the employee case study it is maintained as 0.000001 as the sample values are in hundred to thousand range and it runs over all the samples before it updates particular sample and hence for employee case study it is maintained as 0.000001 $i.e.,$ ${{\eta}_{b} < {\eta}_{s}}$.
\vskip 2mm
{\bf Weight Vector}:The weight vector is a reduction factor to minimize the first stage prediction result. The gradient descent methods runs over several iterations and in every iteration the weight vector is reduced by certain factor until local minimum is reached.
 The weight vector is defined as follows,
\begin{equation}
W_i  = \{ x |x\in R^+,0\leq i < n \}	
\end{equation}
Any $i^{th}$ value of weight vector is an positive numbers. For current implementation every element of weight vector is initialized to 1 and it is reduced by small factor in every iteration.
\vskip 2mm
{\bf UN Partitioned Data Records}:
The data set is a collection of data records where every record is an information about individual object and defined as follows,\\
$X_i$  =\{ ${Ob}^{\rightarrow}_i   |{Ob}^{\rightarrow}$=
\begin{equation}
\{ value1, value2..m \}, ~0\leq i <n \}
\end{equation}
Any $i^{th}$ value of $X$ is a object of m attributes. The entire table structure is of $n\times m$ dimension.
\vskip 2mm
{\bf First Stage Vector}:
This is the result of first stage prediction and is defined as,
\begin{equation}
f_i =\left\lbrace{ x |x\in R^+, 0\leq i<n} \right\rbrace
\end{equation}
Any $i^{th}$  value of weight vector is an positive integer. The prediction is done for every object using RDF metadata and hence the magnitude of $f$  $i.e.,$ $\vert f\vert$ is $n$.
\vskip 2mm
{\bf Second Stage Prediction Function}:
In second stage the gradient descent methods are applied on $f^{\rightarrow}$ (first stage prediction result) by multiplying weight vector in multiple iterations. In every iteration the prediction function is minimized using updated vector. The prediction function is defined as,
\begin{equation}
F(w^{\rightarrow},f^{\rightarrow})=p^{\rightarrow}=w^{\rightarrow 2}* {f^ \rightarrow}
\end{equation}
The prediction function is a square of weight vector multiplied with the first stage prediction function. In order to minimize the function in fine granules without the data loss the quadratic factor is used.  The Eq. (4) can also be defined as,
\begin{equation}
\left\lbrace {p_i={w_i}^2*f_i|0 \leq i<n} \right\rbrace
\end{equation}
Every $i^{th}$ value is predicted as the square of corresponding $i^{th}$ element of weight vector and first stage prediction vector.
\vskip 2mm
{\bf Updation of Weight Vector}:
There are two variations in updating weight vector depending on the type of gradient descent method. The Stochastic gradient approach updates sample as soon as it is encountered it unlike Batch gradient which runs over all samples before it updates any individual sample.
According to Stochastic Gradient Method any $i^{th}$ element of weight vector is updated as per the Eq. (6)
\begin{equation}
{w_i}={w_i}-{\eta}s \bigtriangledown F(w_i,f_i)
\end{equation}
Similarly the Batch Gradient Method updates the weight vector by considering all the elements of weight vector before it updates particular element and hence it is defined as per Eq. (7).
\begin{equation}
w_i=w_i-{\eta}{b_j} \sum\limits_{j=0}^{n-1} \bigtriangledown F(w_j,f_j)
\end{equation}
The $\bigtriangledown F (w_j,f_j)$ is a differential of second stage prediction function w.r.t $w$ (weight vector) and hence it is defined as,
\begin{equation}
\bigtriangledown F({w^ \rightarrow},{f^ \rightarrow})= 2 * w^{\rightarrow}*f^{\rightarrow}
\end{equation}
In every iteration the next state of weight vector is updated based on the previous value. When $ \bigtriangledown F
({w^ \rightarrow},{f^ \rightarrow})$ becomes zero the $ \bigtriangledown F
({w^ \rightarrow},{f^ \rightarrow})$ reaches minimum value and hence learning stops as there is no value change happens from previous to next step.
\vskip 2mm
{\bf Expected Vector}:
The Expected vector is a collection of prediction values for every data record and is defined as,\\
\begin{equation}
E_i  =\{ \sum\limits_{j=0}^{m} X_{i,j}  | 0\leq i<n,0\leq j<m \}
\end{equation}
\vskip 2mm
{\bf Learning Stoppage and Expectation Probability}:
The learning stoppage is the minimum probability that decides the termination point of learning process. The expectation probability is the ratio of least square value of prediction to the expected outcome and is defined as per the Eq. (10).
\begin{equation}
ep= \sum\limits_{i=0}^{n} {{p_i}^2}/2 \div \sum\limits_{i=0}^{n} {{E_i}^2}/2
\end{equation}
The learning process stops when expectation probability hits learning stoppage $i.e.,$ when $ep\lambda$.
\section{ALGORITHMS}
The entire learning process is driven by semantic web based two stage prediction process. Each communicating party generates the required RDF metadata for their respective partitioned data. The first stage prediction process starts with each communicating parties exchanging the RDF metadata for their respective unknown values. The RDF metadata provides high level disguised information that allows communicating parties to infer upper bound values for the unknown attributes.
\vskip 2mm
Once the unknown values are inferred they are processed with known data in second stage prediction process to reach the approximation to the expected vector. Since the first stage prediction operates on upper bound values the gradient descents methods are applied in second stage prediction process to minimize the prediction vector in negative steepest descent until learning stoppage is reached.  The data is partitioned vertically where every communication party has same set of records but with varying attributes.
\begin{table}[ht!]
\begin{center}
\caption{Algorithm for generating RDF Model}
\begin{tabular}{|l|}
\hline
{\bf Input:}\\
1. Ontology\_mod ; common Ontology \\
\indent\hspace{.1in}model used by all parties\\
2. Xa ; $Alice$ partitioned data\\
3. Xb ; $Bob$ partitioned data\\
4. df ; Disguising factor is a factor by\\
\indent\hspace{.1in} which the original data is generalized\\
{\bf Output:}\\
1. RDF\_A ;$Alice$ Department RDF for $Bob$\\
2. RDF\_B ;$Bob$ department RDF for $Alice$\\
{\bf Process:} \\
1. data\_set=Xa;\\
2. attributes=list of attributes of data\_set\\
\indent\hspace{.1in} and first element is always record identifier.\\
3. RDF\_Mod=RDF\_A\\
{\bf do} \\
\indent\hspace{.1in} R\_i= data\_set\{ 0 \};\\
{\bf do}\\
\indent\hspace{.1in}$max\_attr_j$ = get the max value of\\
\indent\hspace{.1in} $j^{th}$ attribute depending on the \\
\indent\hspace{.1in} record criteria.\\
\indent\hspace{.1in} $Rel_j$ = find the relation from \\
\indent\hspace{.1in} $Ontology\_mod$ corresponding to the \\
\indent\hspace{.1in} $j^{th}$ attribute.Val\_j=max\_attr\_{j}+ df ; \\
\indent\hspace{.1in}generalize the value using disguising factor.\\
\indent\hspace{.1in}$RDF < R_i, Rel_j, Val_j > $ = form RDF \\
\indent\hspace{.1in}triple with subject as record identifier ,\\
\indent\hspace{.1in}object as  $Val_j$ and $Rel_j$ as predicate \\
\indent\hspace{.1in}defining the context between subject .\\
\indent\hspace{.1in}and Object. $RDF\_Mod=RDF\_Mod \cap $\\
\indent\hspace{.1in}$RDF < R_i,{Rel}_j,{Val}_j> $\\
    {\bf done}\\
    {\bf done}\\
 5. Repeat the above process $i.e.,$ step 4 \\
\indent\hspace{.1in}for $RDF\_Mod$ set to $RDF\_B$.\\
{\bf done} \\
\hline

\end{tabular}
\end{center}
\end{table}
The algorithm in Table 2  is used for generating RDF Meta data model for the contents of data records. Every data record has certain criteria with which most generalized information is retrieved. For employee record, one such criteria is a type of employee/category of employee. The category in employee context defines the designation or job position like Team Lead, Project Manager $etc..$ Depending on the category the generic information of salary components are retrieved. The above algorithm finds the maximum value of every salary component(PF, Gratuity...) depending on class of employee and hence RDF record is created for every data record with upper bound values for every attribute as generic information. The generic information is further disguised using disguising factor and hence privacy of every data record is retained. Every RDF triples are defined using single ontology model.
\begin{table}[ht!]
\begin{center}
\caption{Algorithm for First Stage Prediction Process}
\begin{tabular}{|l|}
\hline
{\bf Input:}\\
 RDF\_A, RDF\_B \\
{\bf Output:} \\
  $Af^{\rightarrow},Bf^{\rightarrow}$ \\
{\bf Process:} \\
1. $Alice$ initiates communication using \\
\indent\hspace{.2in}CON\_INIT(Connection Initialization\\
\indent\hspace{.2in}Segment)and $Bob$ responds with \\
\indent\hspace{.2in}CON\_INIT\_ACK as acknowledgement\\
\indent\hspace{.2in}
for CON\_INIT.\\
2. $Alice$ request for location of\\
\indent\hspace{.2in}RDF\_B of $Bob$ using REQUEST\\
\indent\hspace{.2in}RDF\_B(Request packet of the \\
\indent\hspace{.2in}format REQUEST followed by\\
\indent\hspace{.2in}RDF  filename) and $Bob$ responds\\
\indent\hspace{.2in}with RESPONSE RDF\_B\_URL($Bob$ \\
\indent\hspace{.2in}responds with the location of RDF)\\
\indent\hspace{.2in}and $Bob$ also piggy back the response\\
\indent\hspace{.2in}with REQUEST RDF\_A(Request\\
\indent\hspace{.2in}for $Alice$ RDF location).\\
3. $Alice$ responds with the location\\
\indent\hspace{.2in}of RDF\_A with the packet RESPONSE\\
\indent\hspace{.2in}RDF\_A\_URL and piggyback the \\
\indent\hspace{.2in}termination request using CON\_TERM.\\
4. $Bob$ responds with CON\_TERM\_ACK as\\
\indent\hspace{.2in}acknowledgement for closing connection.\\
5. $Alice$ reads the Meta data from\\
\indent\hspace{.2in}RDF\_B\_URL and predicts the\\
\indent\hspace{.2in}data for unknown attributes.\\
6. $Bob$ reads the Meta Data from\\
\indent\hspace{.2in}RDF\_A\_URL and predicts the\\
\indent\hspace{.2in}data for unknown attributes.\\
7. $Alice$ and $Bob$ produces \\
\indent\hspace{.2in}$Af^{\rightarrow},Bf^{\rightarrow}$ first stage\\
\indent\hspace{.2in}vectors by summing up the values \\
\indent\hspace{.2in}of known attributes with the\\
\indent\hspace{.2in}unknown values interpreted \\
\indent\hspace{.2in}from RDF model for every record.\\

\hline
\end{tabular}
\end{center}
\end{table}
The algorithm in Table 3 is for first stage prediction. In first stage prediction the $Alice$ and $Bob$ exchanges the RDF model containing disguised abstract information. The abstract information is more generic, like for every employee the values are generalized by finding the maximum value of corresponding attributes based on the record criteria $i.e.,$ type of employee and further the generalized information is disguised using disguising factor. For Example basic salary of any employee who is team lead is updated with maximum of basic salaries of all the employees belonging to team lead class added with disguising factor forming the more generic information.
\begin{table}
\begin{center}
\caption{Algorithm for Second Stage Prediction Process}
\begin{tabular}{|l|}
\hline
{\bf Input:}\\
1. $Af^{\rightarrow},Bf^{\rightarrow}$;\textit{Alice}and \textit{Bob}\\
\indent\hspace{.2in}first stage prediction vectors.\\
2. $w_a^{\rightarrow},w_b^{\rightarrow}$;\textit{Alice} and \textit{Bob}\\
\indent\hspace{.2in}unit weight vectors.\\
3. $E^{\rightarrow}$;  Expected Vector.\\
4. GDTYPE : Type of Gradient Descent\\
\indent\hspace{.2in}Method.\\
\indent\hspace{.2in}Values are either Stochastic or Batch.\\
{\bf Output:}
\indent\hspace{.2in}$p^{\rightarrow}$ ; Second stage vectors as \\
\indent\hspace{.2in}a result of gradient descent \\
\indent\hspace{.2in}methods application where\\
{\bf Process:} \\
1. \textit{Alice} performs $ AP^{\rightarrow} ={w_a}^{\rightarrow 2}*Af^{\rightarrow}$ and\\
\indent\hspace{.2in}\textit{Bob} performs $BP^{\rightarrow}={w_b}^{\rightarrow 2} * Bf^{\rightarrow}$\\
2. \textit{Alice} requests for $AP^{\rightarrow}$ and \textit{Bob}\\
\indent\hspace{.2in}requests for $BP^{\rightarrow}$\\
3. After the successful exchange ,\\
\indent\hspace{.2in}\textit{Alice} and \textit{Bob} calculates\\
\indent\hspace{.2in}$p^{\rightarrow}$ combinedly as = $(AP^{\rightarrow} + BP^{\rightarrow})/2$\\
4. Calculate Expectation Probability ($ep$)\\
\indent\hspace{.2in}according to  equation(6)\\
5.  {\bf if}  $ep\lambda$     {\bf then} \\
\indent\hspace{.2in}{\bf go to} step 6.\\
{\bf else}\\
{\bf if} $GD\underbar TYPE$==”Stochastic”   {\bf then}\\
update the weight Vector\\
$({w_a}^{\rightarrow} ,{w_b}^{\rightarrow} )$ as per  equation(4).\\
           {\bf else}\\
update the weight vector\\
$({w_a}^{\rightarrow} ,{w_b}^{\rightarrow} )$  as per   equation(5).\\
           {\bf end if}\\
         {\bf go to} step 1.\\
     {\bf end if}\\
6. {\bf Stop} the process.\\
\hline

\end{tabular} \\
\end{center}
\end{table}
\vskip 2mm
In First stage prediction process the $Alice$ and $Bob$ mutually exchanges the location of generated RDF data and individually the RDF data is process to find the values of unknown attributes. The first stage prediction vector is produced by summing up the values of known and interpreted values of unknown attributes. The communication parties follows the specific protocol for exchanging the required messages. The protocol has following message segments used,\\
\textbf{CON\_INIT:} Connection initialization segment. The is sent by any communication party to trigger the communication.\\
\textbf{CON\_INIT\_ACK: }Acknowledgement to CONINIT. This is sent by the receiver.\\
\textbf{REQUEST Message: }The Message is a string which represents the request information. In first stage prediction this represents the filename of RDF required. In second stage, message is a name of the vector expected from other communication party.\\
\textbf{RESPONSE Message: }This is the response message for request message. The Message in this context is list of strings separated by a delimiter $\vert$. In first stage process it is of the form RDF\_FILE\_NAME$\vert$ RDF\_URL.Similarly in second stage  it is VECTOR\_NAME$\vert$ JSON(VECTOR\_NAME).\\
Where JSON(VECTOR\_NAME)  is a JSON representation of Vector.\\
\textbf{CON\_TERM:} Request to terminate communication.\\
\textbf{CON\_TERM\_ACK: }Acknowledgement for CON\_INIT.\\
Both communication parties has shared secret key and the messages are encrypted using DES algorithm. All the protocol segments are encrypted and the cipher is exchanged between the communication parties, hence the secure communication adds further privacy to the data.
\vskip 2mm
The Table 4 is an algorithm for second stage prediction. This algorithm takes first stage result, weight vectors, expected vector and Gradient descent type as inputs. It applies the prediction function as shown in step 1 to find the second stage prediction vector. The predicted vector is normalized/fine tuned using gradient descent methods in the direction of negative steepest descent until expectation probability reaches learning stoppage. The algorithm applies only one gradient descent method at a time. This is an iterative algorithm, where in every iteration the weight vector is reduced and prediction function is applied until expectation probability becomes less than or equal to learning stoppage.
\vskip 2mm
As shown in step 5 the weight vectors are updated based on the type of gradient descent method. If the type is stochastic it updates individual element at a time as per Eq. (6) else if the type is batch gradient descent then average of weight vector values are calculated before it updates any element of weight vector as per Eq. (7). The algorithm runs with single gradient descent method at a time. The second stage algorithm uses the same protocol segments as used in first stage prediction for exchange of messages and all the messages are encrypted before they are exchanged.
\vskip 2mm
The sub graphs are processed to from all possible RDF-triplets which are submitted to the database to retrieve URL set. Since there is a probability that RDF-triplets are repeated in multiple web pages the final URL set is obtained by the intersection of URL sets of all RDF-triplets matching the user query as explained in the algorithm of Table 5. 
\section{SYSTEM ARCHITECTURE}
As shown in Figure 2. The data records are vertically partitioned between the two processes ($Alice$ and $Bob$).  Vertical partitioning means that every process has same set of records with disjoint attributes set similarly Horizontal partitioning means that every process has disjoint set of records but with the same attributes. Complete prediction process is divided into two stages.
\begin{figure}[ht!]
\centerline{\includegraphics [width=8cm]{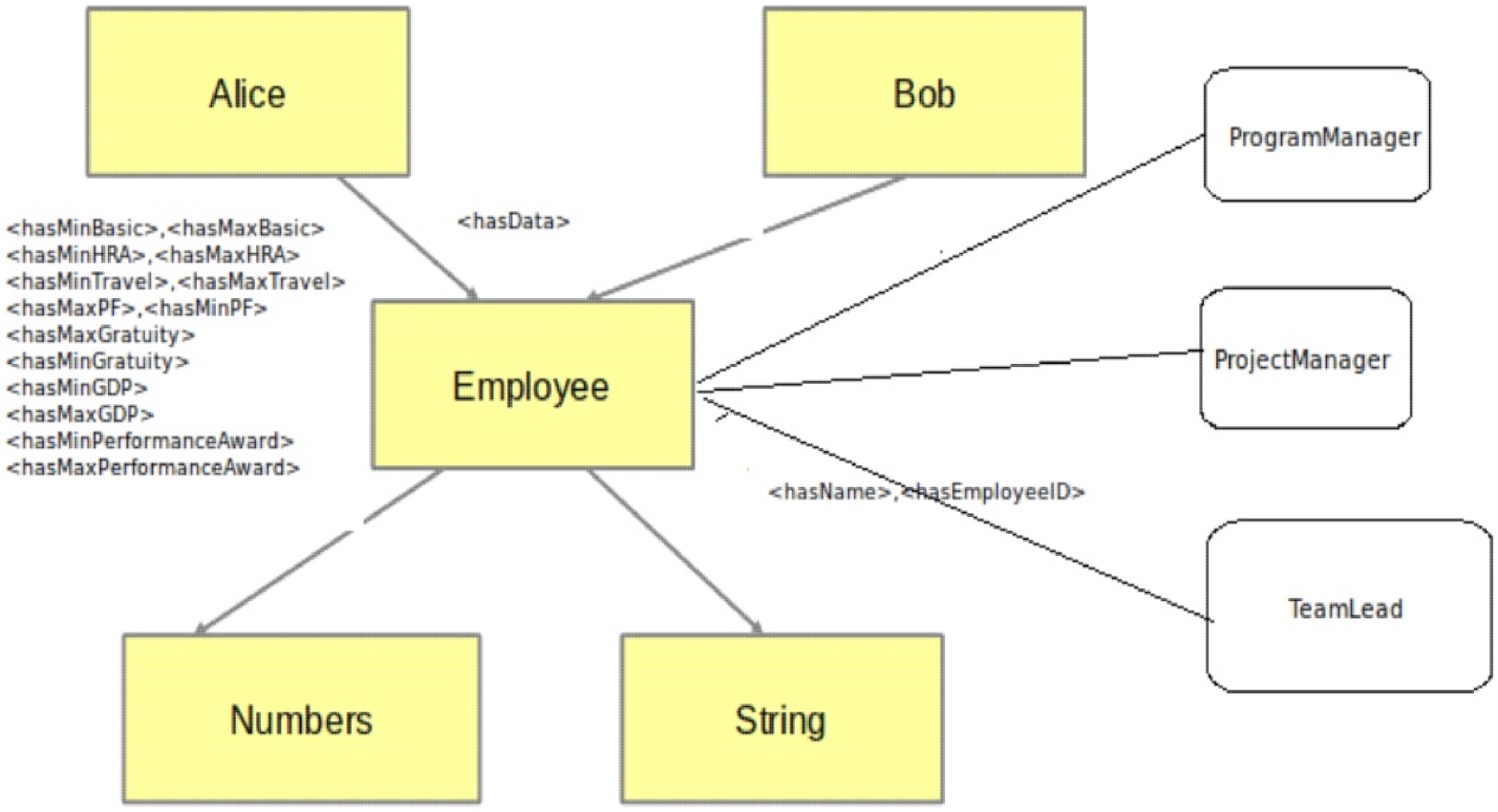}}
\caption{Ontology Model for Employee Domain}\label{Fig1}
\end{figure}
\subsection{Ontology Model}
The Ontology defines concepts and relations between the concepts. Ontology is domain specific and it is generated based on the database schema. Ontology defines concept for every table in the schema and relation between the tables are defined as Object and Data type properties. The object property describes the object from other object and the object being described is referred as subject. The data type property describes the subject using the textual information for example the email of $Bob$ can be described with the relation $<$hasEmail$>$ between the person and String. Where $Bob$ belongs to the concept person and his email say $bob@example.com$ is conceptualized as a String. The ontology model gives the specification for which the multiple RDF’s adhered to.

\subsection{RDF Metadata and First Stage Prediction}
The RDF represents a metadata for the corresponding partitioned data. The metadata is high level generalized information. The generalized information is further disguised to increase the generality using the disguising factor. The RDF is represented as collection of triplets which are defined according to the corresponding domain ontology. The $Alice$ and $Bob$ exchange’s the RDF information mutually before the actual prediction process. The data inferred from the RDF is the result of first stage prediction process. The result of first stage prediction process defines the model for second stage prediction.
\vskip 2mm
The model in the machine learning context defines the knowledge extracted from the training data with aid of past experience but here the new approach is used to define the rules/model/knowledge from the actual data. The RDF content is used as semantics to define the model instead of the training data. Since the semantics is more relevant or nearest to the actual data but not data itself and hence the privacy is preserved. The result of the first stage prediction process in this context is the regression model.The Fig 1 shows the ontology model for Employee domain. The ontology model is a definition of concepts and relations for generalization. The ontology model defines $<$haxMax$>$ and  $<$hasMin$>$ relations between the Employee and Numbers domain. The $<$haxMax$>$ defines the maximum value of attribute among the records belonging to specific category like Project Manager, Team Lead and Program Manager. For example $<$hasMaxBasic$>$ represents maximum value of basic component of salary and this is calculated for every category to create generalized information. Similarly $<$hasMinBasic$>$ represents minimum value of basic component of salary. All the attributes are represented as datatype properties except $<$hasData$>$ which represents the partitioned data of Employee database. The RDF generator uses the relations defined in ontology model to represents the most generalized information as semantics for known attributes. The sample RDF generated is shown in Table 5.
\begin{table}[ht!]
\begin{center}
\caption{Sample RDF of $Alice$  for $Bob$}
\begin{tabular}{|l|}
\hline
{\footnotesize $<$ rdf:RDF xmlns:j.0="http://www.ppgd.com/"}\\
\hspace{15pt}{\footnotesize xmlns:rdf="http://www.w3.org/1999/02/22-}\\
rdf-syntax-ns\#" $>$\\
\footnotesize   $<$ rdf:Description rdf:about \\
="http://www.SkumarSolutions.com/ID10" $>$\\
\hspace{15pt}{\footnotesize $<$j.0:hasMinflat$>$37$<$/j.0:hasMinflat$>$}\\
\hspace{15pt}{\footnotesize $<$j.0:hasMinTravel$>$38$<$/j.0:hasMinTravel$>$}\\
\hspace{15pt}{\footnotesize $<$j.0:hasMaxTravel$>$55$<$/j.0:hasMaxTravel$>$}\\
\hspace{15pt}{\footnotesize $<$j.0:hasMinBasic$>$20$<$/j.0:hasMinBasic$>$}\\
\hspace{15pt}{\footnotesize $<$j.0:hasMaxflat$>$45$<$/j.0:hasMaxflat$>$}\\
\hspace{15pt}{\footnotesize $<$j.0:hasMaxBasic$>$30$<$/j.0:hasMaxBasic$>$}\\
\hspace{15pt}{\footnotesize $<$n.0:hasMinHRA$>$30$<$/j.0:hasMinHRA$>$}\\
\hspace{15pt}{\footnotesize $<$j.0:hasMaxHRA$>$42$<$/j.0:hasMaxHRA$>$}\\
\hspace{15pt}{\footnotesize $<$j.0:hasName$>$reva123$<$/j.0:hasName$>$}\\
{\footnotesize $<$/rdf:Description$>$}\\
\hline
\end{tabular}
\end{center}
\end{table}

\begin{figure*}[ht!]
\centerline{\includegraphics [width=10cm]{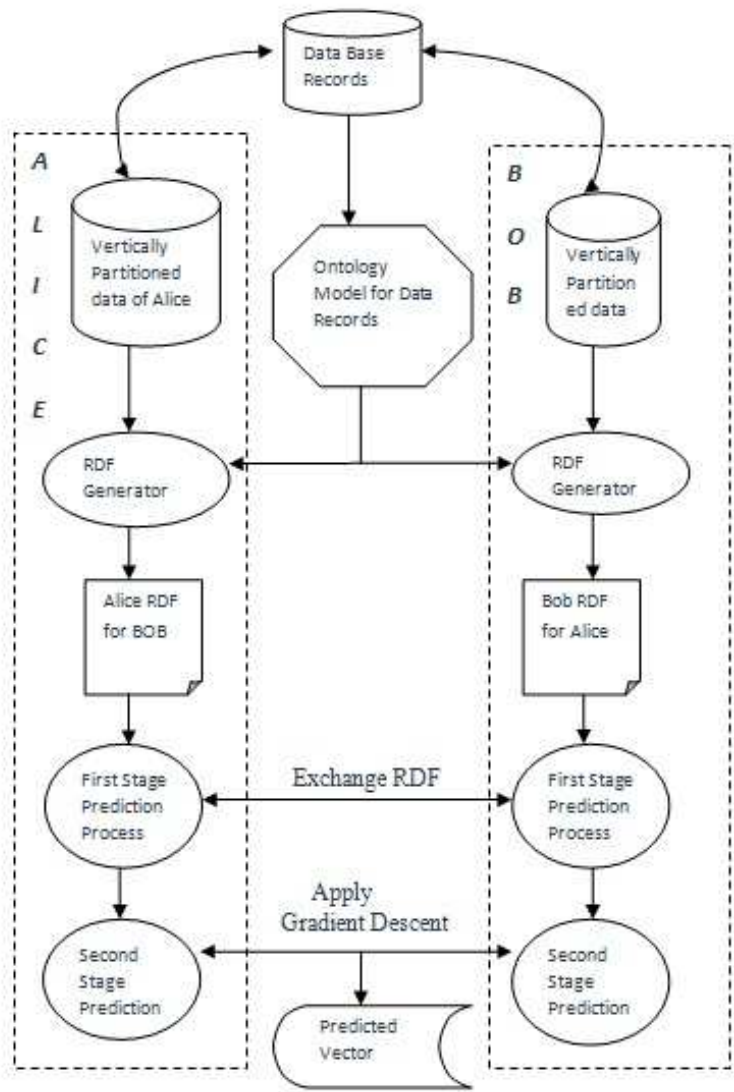}}
\caption{Components in TPPGD}\label{Fig3}
\end{figure*}

The RDF Metadata  is a high level generalized information disguised with certain disguising factor to preserve the privacy of data. The disguising factor is not disclosed between the communicating parties, each of them can use his/her own disguising factor. $Alice$ and $Bob$ combinedly defines the model from the deceived data.
\subsection{RDF Generator}
The RDF generator is a vital component which is responsible for generating required RDF semantics for the vertically partitioned data. This component runs over every record and creates respective generalized data depending on the category of record. The generalized information is deceived to other communicating party with the aggregation of disguising factor. The disguised information is used to infer the values of unknown attributes of records and finally a regression model is defined as a result.
\subsection{Second Stage prediction}
In second stage prediction the regression model is used as input for further optimization. The coefficients of regression model are optimized using gradient descent methods. The batch/stochastic gradient descents are applied in negative steepest descent until the threshold is reached. The threshold indicates the fine granular value below which the further optimization is not required and hence it is a stoppage point for the prediction. The stochastic gradient descent is applied to reduce weight vector by considering single element at a time where as batch gradient runs on all elements before it updates particular element of weight vector. In every iteration the weight vector is updated and the predicted vector after every iteration is exchanged securely. Any data exchanged between the communication parties are encrypted using common secret key with the known cryptographic algorithm. This adds the security further to the privacy preserved data.
\section{EXPERIMENTAL RESULTS}
The experiment is carried out on the employee database for calculation of salary. The scenario is $Alice$ and $Bob$ heads payroll department with distributed database. The employee table is vertically partitioned in such a way that attributes forming the  salary component is distributed across $Alice$ and $Bob$ department. The identified attributes of salary components are Basic, HRA, flat, Travel, PF, Gratuity, GD and Performance Award. After partitioning, the $Alice$ has \{ Basic, HRA, flat, Travel \} and $Bob$ has \{ PF, Gratuity, GD and Performance Award \} as shown in Figure 3.
\begin{figure}[ht!]
\begin{center}
{\footnotesize EmpID $\vert$ name$\vert$ Basic$\vert$ HRA$\vert$
flat$\vert$ Travel$\vert$ PF$\vert{}$ Gratuity $\vert$ GDP ~$\vert$PerformanceAward
$\vert{}$Category$\vert{}$
\includegraphics[width=60pt]{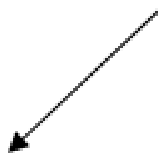}\includegraphics[width=60pt]{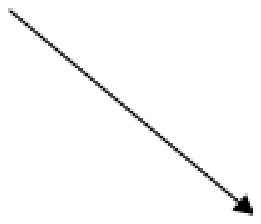}
\begin{flushleft}
 ~(Alice):~
$\vert$ EmpID$\vert$Basic$\vert$ \\
$\vert$ HRA$\vert$flat$\vert$ \\
$\vert$ Cravel$\vert$Category$\vert$ \\
\end{flushleft}

\begin{flushright}
~(Bob):~
$\vert$ EmpID$\vert$PF$\vert$Gratuity$\vert$ \\
$\vert$ GDP$\vert$PerformanceAward$\vert$ \\
$\vert$ Category$\vert$  \\
\end{flushright}
}
\end{center}
\begin{center}
\caption{Vertically Partitioned Data}\label{Fig2}

\end{center}
\end{figure}
The Employee Id and Category is used in both departments for creating RDF semantics information. The category in this scenario indicates the designation of employee which is used as key attribute for generalizing the data. The generalization is performed by taking max value of known attributes depending on the category. For example, from the existing database content the maximum of Basic, PF $etc.,$ is determined for every category like teamLead, projectManager and programManager and further it is wrapped with disguising factor. $Alice$ and $Bob$ performs the above operation with their corresponding known attributes. The disguised information is serialized as RDF data and exchanged. Since the known attributes of $Alice$ is unknown attributes for $Bob$ and vice versa is true and therefore $Alice$/$Bob$ mutually finds the values of unknown attributes using RDF semantics. The first stage prediction is performed by the summation of values of unknown attributes inferred from the RDF and known attribute values for every employee record. This results in single salary vector defining the model for second stage prediction process at each department. Experiment is carried out with the disguising factor set to 10\$ during the first stage prediction process. The identified  maximum values  are disguised by adding 10\$ to the amount.
\vskip 2mm
$Alice$ and $Bob$ performs the optimization of regression model obtained as a result of first stage prediction process. The optimization is achieved by applying gradient descent methods on the predicted output. In gradient descent the weight vector is used as a coefficient for prediction function as defined in Eq. (5) and the weight vector is optimized as per Eq. (6) during stochastic gradient and as per Eq. (7) during batch gradient descent. In stochastic gradient descent the optimization happens in with less factor in each iteration as it update one element of weight vector at a time whereas in batch gradient the optimization happens with high factor.
\begin{figure}[ht!]
\centerline{\includegraphics [width=8cm]{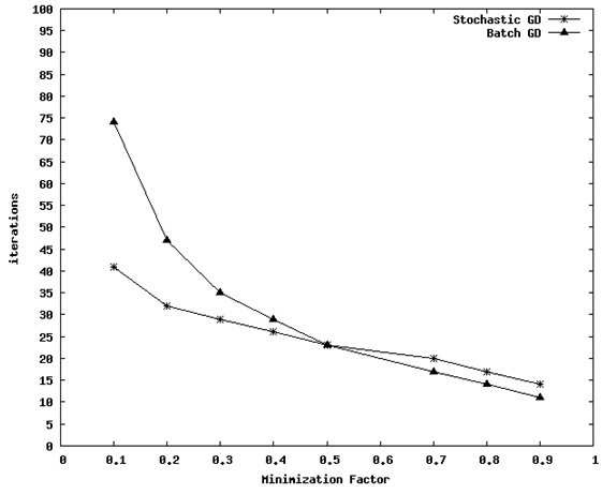}}
\caption{Batch vs Stochastic Gradient with respect to no of Iterations.}\label{Fig4}
\end{figure}
\\
\
\hspace{15pt}As a result of which the batch gradient descent takes less number of iterations to predict output for high minimization factor/learning stoppage as shown in Figure 4, which indicates for high minimization factor/learning stoppage the batch gradient descent method takes more iterations until switching point of 0.5 minimization factor after which the behavior of batch and stochastic remains stagnant. The stochastic gradient takes less number of iterations for prediction after 0.5 $i.e.,$ for lower values of minimization factor.
\begin{figure}[ht!]
\centerline{\includegraphics [width=8cm]{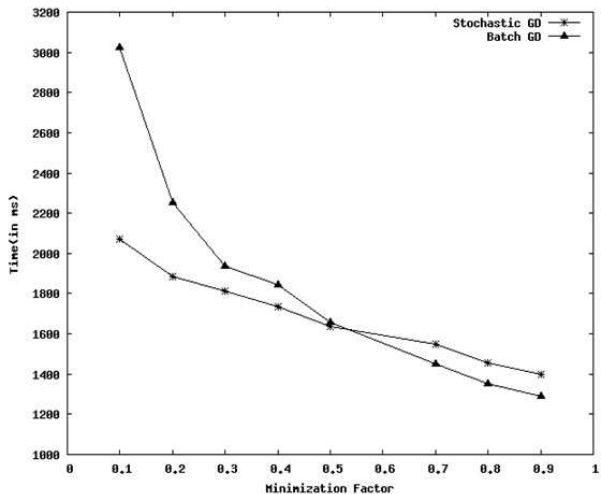}}
\caption{Batch vs Stochastic Gradient with respect to Execution Time.}\label{Fig5}
\end{figure}
Similarly the execution time for batch gradient is high as number of iterations are more for prediction up to switching point as shown in Figure 5. After switching point the stochastic the batch gradient takes more time than stochastic for lower values of minimization factor. The experiment is carried out by varying learning stoppage from high value to low and behavior of stochastic and batch gradient descents are analyzed with respect to number of iterations and time required to perform prediction.
\vskip 2mm
The DES algorithm is used as cryptographic algorithm to retain confidentiality of data exchanged between $Alice$ and $Bob$. The experiment is simulated with raw sockets and implemented using JAVA. The $Bob$ socket is used as server socket waiting for $Alice$ to initiate communication. In First stage prediction the RDF location is exchanged securely and in second stage the regression model which obtained as result from first stage prediction is exchanged securely between $Alice$ and $Bob$.

\section{CONCLUSION}
In this paper a new approach is proposed to retain the privacy of data with the combination of generalization and bamboozling. The bamboozling is the process where the second level privacy is achieved by disguising the generalized information with certain factor to deceive the other communication parties. The complete generalization and bamboozling process happens as first stage prediction process to define the regression model which is used as input for second stage prediction for optimization and predict accurate result.
\vskip 2mm
The experimental results shows the behaviour of batch and stochastic process with respect to the number of iterations and execution time required for prediction process. It is shown that batch gradient descent takes long time and more iterations for higher values of minimization factor than stochastic method. The solution can be enhanced by adding the suppression technique along with generalization and bamboozling processes to further strengthen the privacy of data.
\small
\balance

\noindent{\includegraphics[width=1in,height=1.7in,clip,keepaspectratio]{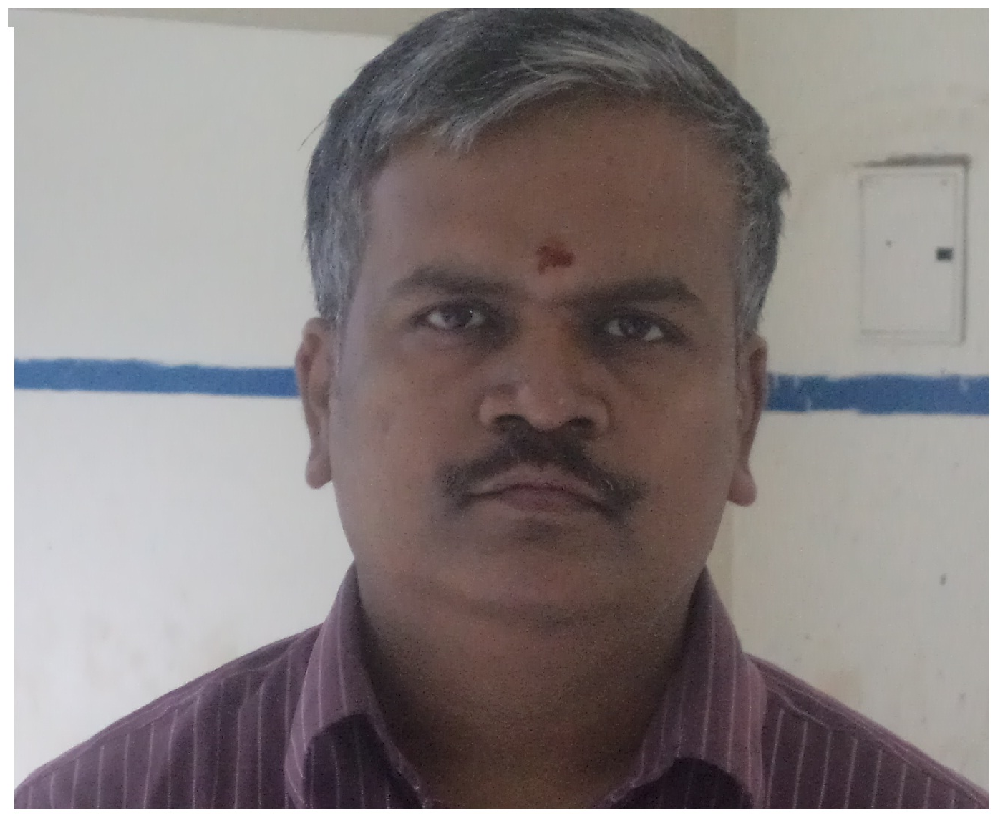}}
\begin{minipage}[b][1in][c]{1.8in}
{\centering{\bf {Kumaraswamy S}} is currently working as an Assistant Professor in the Department of Computer Science and Engineering, KNS Institute of Technology, Bangalore, India. He obtained }
\end{minipage} 
his  Bachelor of Engineering from SiddaGanga Institute of Technology, Tumkur. Bangalore University, Bangalore. He is presently pursuing his Ph.D programme in the area of privacy management in databases in Bangalore University. His research interest is in the area of Data Mining, Web Mining and Semantic Web.\\\\
\noindent{\includegraphics[width=1in,height=1.7in,clip,keepaspectratio]{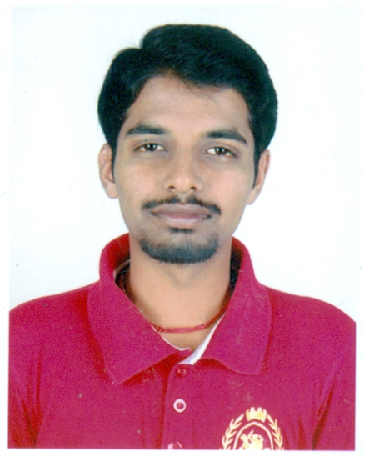}}
\begin{minipage}[b][1in][c]{1.8in}
{\centering{\bf {Srikanth P L}} received his Master's degree from the Department Computer Science and Engineering, University Visvesvaraya College of Engineering, Bangalore University, Bangalore. 
His research interest is in the area of Web Technology, Se-}\\  
\end{minipage}
mantic Web and Cloud Computing. \\\\

\noindent{\includegraphics[width=1in,height=1.7in,clip,keepaspectratio]{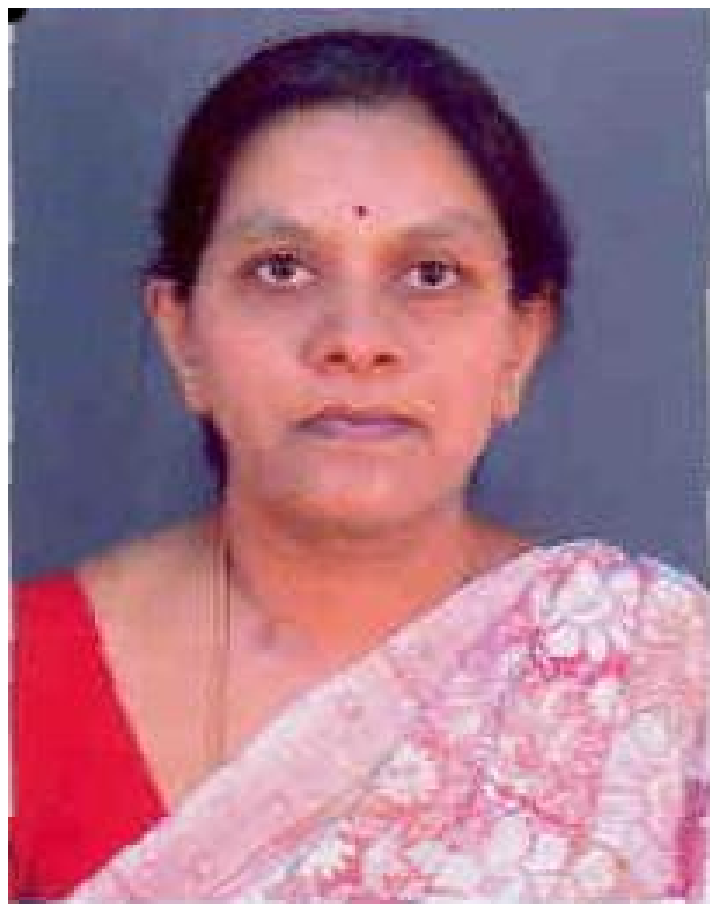}}
\begin{minipage}[b][1in][c]{1.8in}
{\centering{\bf {S H Manjula }} is currently the Chairman, Department of Computer Science and Engineering, University Visvesvaraya College of Engineering, Bangalore University, Bangalore. She obtained her Bachelor of Engineering and Masters Degree in Computer Science and Engineering from }\\\\  
\end{minipage}
University Visvesvaraya College of Engineering. She was awarded Ph.D. in Computer Science from Dr. MGR University, Chennai. Her research interests are in the field of Wireless Sensor Networks and Data mining. \\\\

\noindent{\includegraphics[width=1in,height=1.7in,clip,keepaspectratio]{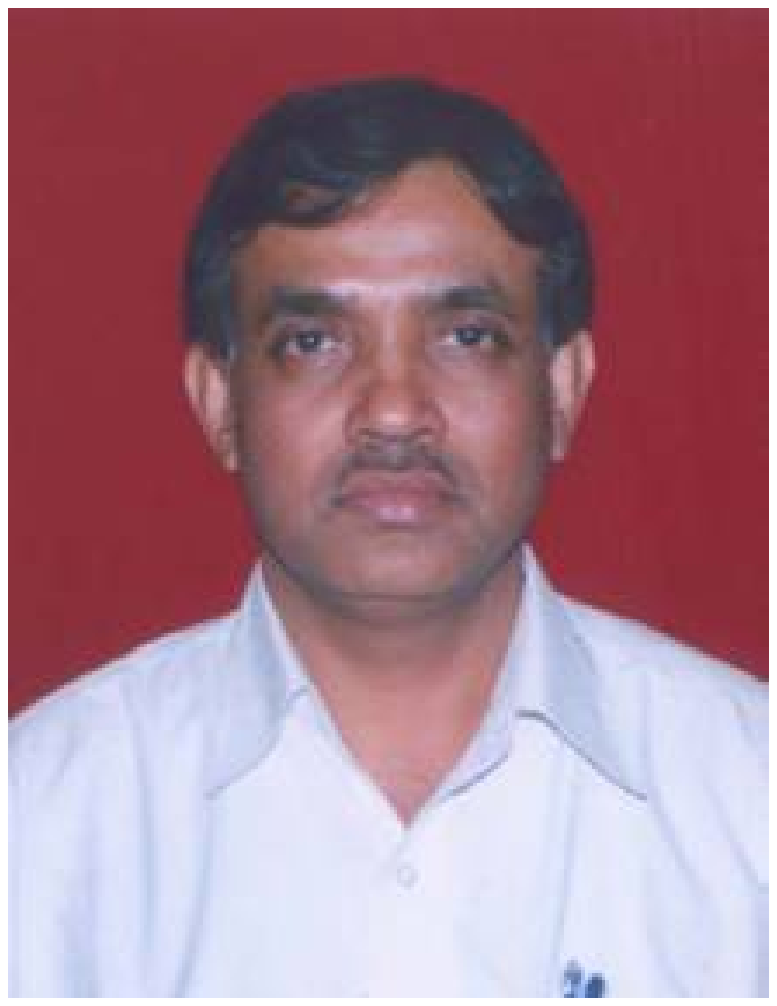}}
\begin{minipage}[b][1in][c]{1.8in}
{\centering{\bf {K R Venugopal}} is currently the Principal, University Visvesvaraya College of Engineering, Bangalore University, Bangalore. He obtained his Bachelor of Engineering from University Visvesvaraya College of Engineering. He received his Masters degree in Computer Science and }\\\\
\end{minipage}
 Automation from Indian Institute of Science Bangalore. He was awarded Ph.D in Economics from Bangalore University and Ph.D in Computer Science from Indian Institute of Technology, Madras. He has a distinguished academic career and has degrees in Electronics, Economics, Law, Business Finance, Public Relations, Communications, Industrial Relations, Computer Science and Journalism. He has authored 39 books on Computer Science and Economics, which include Petrodollar and the World Economy, C Aptitude, Mastering C, Microprocessor Programming, Mastering C++ and Digital Circuits and Systems $etc.$. During his three decades of service at UVCE he has over 350 research papers to his credit. His research interests include Computer Networks, Wireless Sensor Networks, Parallel and Distributed Systems, Digital Signal Processing and Data Mining. \\\\
\noindent{\includegraphics[width=1in,height=1.7in,clip,keepaspectratio]{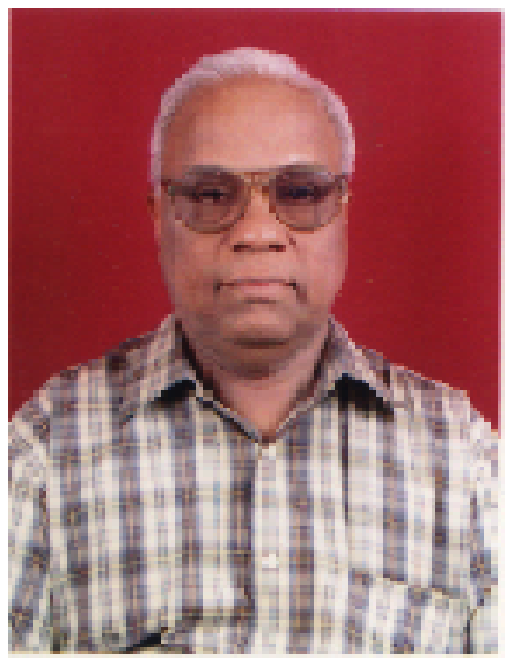}}
\begin{minipage}[b][1in][c]{1.8in}
{\centering{\bf{L M Patnaik }}is currently Honorary Professor, Indian Institute of Science, Bangalore, India. He was a Vice Chancellor, Defense Institute of Advanced Technology, Pune, India and was a Professor since 1986 with the Department of Computer Science and Automation, Indian}\\\\
\end{minipage}
 Institute of Science, Bangalore. During the past 35 years of his service at the Institute he has over 700 research publications in refereed International Journals and Conference Proceedings. He is a Fellow of all the four leading Science and Engineering Academies in India; Fellow of the IEEE and the Academy of Science for the Developing World. He has received twenty national and international awards; notable among them is the IEEE Technical Achievement Award for his significant contributions to High Performance Computing and Soft Computing. His areas of research interest have been Parallel and Distributed Computing, Mobile Computing, CAD for VLSI circuits, Soft Computing and Computational Neuroscience.\\\\
\balance

\begin{thebibliography}{00}
\bibitem{1}
Microsoft research.Database Privacy, {\it http:// research.microsoft.com/en-us/projects/ DatabasePrivacy}.
\bibitem{2}
Latanya Sweeney. {Achieving K-Anonymity Privacy Protection Using Generalization and Suppression,}{ \it In International Journal on Uncertainty, Fuzziness and Knowledge-based Systems}, 10(5): 571-588, 2002.
\bibitem{3}
Gabriel Ghinita, Member, IEEE, Panos Kalnis and Yufei Tao. {Anonymous Publication of Sensitive Transactional Data,} {\it In IEEE Transactions on Knowledge and Data Engineering}, 23(2): 161-174, Feb 2011.
\bibitem{4}
Hopoper N, Saunders J and McHugh L. {The Derived Generalization of Thought Suppression,} {\it Learn Behav}, 38(2): 160-168, 2010.
\bibitem{5}
  Microsoft research. {Database Privacy,} {\it http:// research.microsoft.com/en-us/projects/ DatabasePrivacy}
\bibitem{6}
Shuguo Han, Student Member, IEEE Computer Society, Wee Keong Ng, Member, IEEE   Computer Society, Li Wan and Vincent C S Lee. {Privacy-Preserving Gradient-Descent Methods,} {\it IEEE Transactions on Software Engineering}, 22(6):884-899, June 2010.
\bibitem{7}
Afshar P. {Gradient Descent Optimisation for ILC-based Stochastic Distribution Control,} {\it IEEE International Conference on Control and Automation(ICCA)}, 1134–1139, 2009.
\bibitem{8}
Zhi Ding, Junqiang Hu and Dayou Qian. {On Steepest Descent Adaptation: A Novel Batch Implementation of Blind Equalization Algorithms,} {\it Global Telecommunications   Conference (GLOBECOM 2010)IEEE}, 1-6, 2010.
\bibitem{9}
Gannot S. {Iterative-batch and Sequential Algorithms for Single Microphone Speech Enhancement,} {\it IEEE International Conference on Acoustics, Speech and Signal Processing (ICASSP-97)}, 2:1215-1218, 1997.
\bibitem{10}
Gonzalez A. {A Note on Conjugate Natural Gradient Training of Multilayer Perceptrons,} {\it International Joint Conference on Neural Networks (IJCNN '06)}, 887-891, 2006.

\bibitem{11}
Ningning Jia, E Y Lam. {Stochastic Gradient Descent for Robust Inverse Photomask Synthesis in Optical Lithography,} {\it 17th IEEE International Conference on Image Processing}, 2010.
\bibitem{12}
S Bonnabel. {Stochastic Gradient Descent on Riemannian Manifolds,} {\it IEEE Transactions on Automatic Control}, 58(9): 2217-2229, 2013.
\bibitem{13}
D Beckett. {RDF/XML Syntax Specification (Revised),} {\it http:// www.w3.org/TR/2 004/REC-rdf-syntax-grammar-20040210/}, 1994.
\bibitem{14}
John Hebeler, Matthew fisher, Ryan Blac, Andrew perez-lopez. {Semantic-Web  Programming,} {\it Third Edition, Wiley India pvt.ltd}, 2009.
\bibitem{15}
Stefan Decker, Sergey Melnik, Frank Van Harmelen, Dieter Fensel, Michel Klein, Jeen Broekstra, Michael Erdmann and Ian Horrocks. {The Semantic Web: The Roles of XML and RDF,} {\it IEEE Internet Computing}, pages 63-73, 2000.
\bibitem{16}
Kanishka Bhaduri, Member, IEEE, Mark D Stefanski and Ashok N Srivastava, Senior Member, IEEE. {Privacy Preserving Outlier Detection through Random Nonlinear Data Distortion,} {\it IEEE Transactions on Systems, Man and Cybernetics}, 41(1):260-272, 2011.
\bibitem{17}
Benjamin C. M. Fung, Member, IEEE, Thomas Trojer, Patrick C. K. Hung, Member,IEEE, Li Xiong, Khalil Al-Hussaeni and Rachida Dssouli. {Service-Oriented Architecture  for High-Dimensional Private Data Mashup,} {\it In IEEE Transactions on Services Computing}, 5(3):373-386, 2012.
\bibitem{18}
Jung Yeon Hwang, Sokjoon Lee, Byung-Ho Chung, Hyun Sook Cho and DaeHun Nyang. {Short Group Signatures with Controllable Linkability,} {\it Workshop on   Lightweight Security and Privacy: Devices, Protocols and Applications}, 44-52, March 2011.
\bibitem{19}
ZHU Yu-quan, TANG Yang CHEN Geng. {A Privacy Preserving Algorithm for Mining Distributed Association Rules,} {\it In International Conference on Computer and Management (CAMAN)}, 1-4, May 2011.
\bibitem{20}
Alberto Trombetta, Wei Jiang, Elisa Bertino and Lorenzo Bossi. {Privacy-Preserving Updates to Anonymous and Confidential Databases,} {\it In IEEE Transactions on Dependable and Secure Computing}, 8(4):578-587, July-August 2011.\\

\small
\balance

\end{thebibliography}
\end{document}